\documentclass[11pt]{article}
\usepackage{jheppub}

\usepackage{epsfig}
\usepackage{amssymb}
\usepackage{amsfonts}
\usepackage{amsbsy}
\usepackage{amsmath}
\usepackage{dsfont}
\usepackage{bbm}
\usepackage{upgreek}
\usepackage{amssymb,amscd}
\usepackage{graphicx}
\usepackage{mathrsfs}
\usepackage{amsmath,amsthm}
\usepackage{slashed}
\usepackage{dsfont}
\usepackage{tikz}
\usepackage[utf8]{inputenc}
\makeatletter
\DeclareFontFamily{OMX}{MnSymbolE}{}
\DeclareSymbolFont{MnLargeSymbols}{OMX}{MnSymbolE}{m}{n}
\SetSymbolFont{MnLargeSymbols}{bold}{OMX}{MnSymbolE}{b}{n}
\DeclareFontShape{OMX}{MnSymbolE}{m}{n}{
    <-6>  MnSymbolE5
   <6-7>  MnSymbolE6
   <7-8>  MnSymbolE7
   <8-9>  MnSymbolE8
   <9-10> MnSymbolE9
  <10-12> MnSymbolE10
  <12->   MnSymbolE12
}{}
\DeclareFontShape{OMX}{MnSymbolE}{b}{n}{
    <-6>  MnSymbolE-Bold5
   <6-7>  MnSymbolE-Bold6
   <7-8>  MnSymbolE-Bold7
   <8-9>  MnSymbolE-Bold8
   <9-10> MnSymbolE-Bold9
  <10-12> MnSymbolE-Bold10
  <12->   MnSymbolE-Bold12
}{}

\let\llangle\@undefined
\let\rrangle\@undefined
\DeclareMathDelimiter{\llangle}{\mathopen}%
                     {MnLargeSymbols}{'164}{MnLargeSymbols}{'164}
\DeclareMathDelimiter{\rrangle}{\mathclose}%
                     {MnLargeSymbols}{'171}{MnLargeSymbols}{'171}
\makeatother

\def\be{ \begin{equation} }
\def\ee{ \end{equation}}

\newcommand{\eq}[1]{\begin{align}\begin{split}#1\end{split}\end{align}}

\def\exp{{\rm exp}}

\def\half{\frac{1}{2}}

\def\one{{\hbox{ 1\kern-.8mm l}}}

\def\CA{{\cal A}}

\def\CH {{\cal H}}

\def\CL {{\cal L}}

\def\CN {{\cal N}}

\def\CO {{\cal O}}

\def\CO {{\cal O}}

\def\CH {{\cal H}}

\def\CT {{\cal T}}
\def\CU {{\cal U}}

\def\IC{\mathbb{C}}

\def\ICP{\mathbb{CP}}

\def\IZ{{\mathbb{Z}}}

\def\rmk#1{\bigskip\noindent{\bf Remark} }
\def\cnj#1{\bigskip\noindent{\bf Conjecture:} }

\def\hatG{{\widehat{G}}}

\DeclareMathAlphabet{\mathpzc}{OT1}{pzc}{m}{it}

\def\Tr{ \, \textrm{Tr} \, }

\title{Anomaly Enforced Gaplessness for Background Flux Anomalies and Symmetry Fractionalization} 

\author[a]{T.~Daniel Brennan}
\author[a]{and Aiden Sheckler}

\affiliation[a]{Department of Physics, University of California San Diego,\\
 \textit{9500 Gilman Drive, La Jolla CA 92093-0319, USA}}

\emailAdd{tbrennan@ucsd.edu}
\emailAdd{asheckler@ucsd.edu}

\abstract{Anomalous symmetries are known to strongly constrain the possible IR behavior along any renormalization group (RG) flow. Recently, the extension of the notion of symmetry in QFT has provided new types of anomalies with a corresponding new class of constraints on RG flows. In this paper, we derive the constraints imposed on RG flows from anomalies that can only be activated in the presence of specific background fluxes even though they do not necessarily correspond to a symmetry.
 We show that such anomalies can only be matched by gapped theories that exhibit either spontaneous symmetry breaking or symmetry fractionalization. In addition, we exhibit previously unstudied examples of these flux background anomalies that arise in $4d$ QCD and $4d$ SUSY QCD. 
}

\begin{document}

\maketitle

\section{Introduction}

Symmetry is a powerful tool in the study of quantum field theory. As in the case of classical field theories, symmetries provide strong constraints on quantum dynamics. While symmetries in classical theories lead to foliations of phase space, symmetries in quantum theories lead to a decomposition of the Hilbert space into representations of the symmetry. This decomposition leads to  selection rules for correlation functions,  protection against perturbative and non-perturbative corrections, and can strongly constrain renormalization group flows.

 Recently, the notion of symmetry in QFT has been expanded beyond the framework of group theory to that of category theory. This is the notion of ``generalized'' or ``categorical'' symmetries which provides a new set of tools with which to study QFTs. For a review see \cite{Cordova:2022ruw,Freed:2022qnc,Gaiotto:2014kfa,Schafer-Nameki:2023jdn,Brennan:2023mmt,Bhardwaj:2023kri,Shao:2023gho}. 

 In this paper we will restrict our attention to group-like generalized symmetries and the constraints they impose on the IR behavior of QFTs. These symmetries go under the name of ``higher form'' global symmetries. A higher form global symmetry is labeled by a degree $p\in \IZ_{\geq0}$ and a group $G$ which together are denoted $G^{(p)}$ and is called a $p$-form global symmetry. These symmetries act on $p$-dimensional charged operators via co-dimension-$(p+1)$ topological operators called symmetry defect operators. These topological operators are written as  $U_g(\Sigma)$ where $g\in G^{(p)}$ and $\Sigma$ is a closed co-dimension-$(p+1)$ manifold. These operators obey group multiplication 
 \eq{
 U_{g_1}(\Sigma)\cdot U_{g_2}(\Sigma)=U_{g_1g_2}(\Sigma)~, 
 }
 and act on charged operators via linking. Due to the fact that these operators are topological, all $p$-form global symmetries are abelian for $p>0$. 

The paradigm of the renormalization group flow in QFT, is that any theory  flows in the IR to a scale invariant fixed point which may either be a gapless phase (CFT) or gapped phase (TQFT). The IR phase may additionally either preserve or spontaneously break a global symmetry which is equivalent to the statement of that the vacuum state of the theory either transforms trivially or non-trivially under the global symmetries. This can generally be diagnosed by whether or not a charged operator has a non-trivial vacuum expectation value. 
 
  In addition to whether or not the symmetry is gapped or gapless, we can also differentiate whether or not global symmetries in the IR phase are {fractionalized} \cite{Delmastro:2022pfo,Brennan:2022tyl,Barkeshli:2014cna,Chen:2014wse}. \emph{Symmetry fractionalization} occurs when a $p$-dimensional operator transforms projectively under a $(p-1)$-form global symmetry. Categorically speaking, this implies that co-dimension $(p+1)$ junctions of $(p-1)$-form symmetry defect operators act effectively as $p$-form symmetry defect operators on $p$-dimensional objects (subject to a cocycle condition). 
   A standard example of this is when there is a 0-form global symmetry $G^{(0)}$ with non-trivial $\pi_1(G^{(0)})$ and there exists a line operator that transforms projectively under $G^{(0)}$. This can occur, for example, in gauge theories when a massive charged particle that transforms projectively under $G^{(0)}$ is integrated out. This notion of symmetry fractionalization crucially relies on the fact that we can make a consistent choice of (renormalization) scheme so that the only allowed counter terms are dependent on $G^{(0)}$-background gauge fields which allows us to track whether or not a line operator transforms projectively under $G^{(0)}$.\footnote{Note that it is also consistent to define the global symmetry to be the simply connected cover $\hatG^{(0)}$ where part of the symmetry group acts ineffectively. This allows for schemes that can trivialize any $\hatG^{(0)}$ transformation of a line operator. Such a scheme is consistent, but is not useful as it washes out physically meaningful information. } The projective transformation of a line operator under $G^{(0)}$ is also reflected by the fact that the operator carries a world volume anomaly under shifts of integer lift of obstruction classes $w_2(G^{(0)})\in H^2(BG^{(0)})$ associated with $\pi_1(G^{(0)})$. 
   

As with ``ordinary'' global symmetries, higher form global symmetries provide many powerful tools for studying QFTs and in particular can be used to constrain RG flows. These global symmetries and their anomalies can constrain the behavior of the IR phase of the theory because certain phases are incompatible with different realizations of the symmetry. For example, continuous symmetries cannot be matched by gapped phases where the symmetry is spontaneously broken, whereas for discrete symmetries a spontaneous symmetry breaking phase may be gapped or gappless.

In general, the possible IR phases can be further constrained if any of the global symmetries are anomalous (i.e. if they have an `t Hooft anomaly). The reason is that `t Hooft anomalies must be matched along symmetry preserving RG flows and, as it turns out, many anomalies cannot be matched by symmetry preserving TQFTs. Thus, if a QFT has a symmetry with such an anomaly, it cannot flow to a symmetry preserving gapped phase.  These types of constraints are often referred to as ``Lieb-Schultz-Mattis-type theorems'' or ``anomaly enforced gaplessness'' which have been well studied in the literature \cite{Lieb:1961fr,Wang:2014lca,Wang:2016gqj,Sodemann:2016mib,Wang:2017txt,Kobayashi:2018yuk,Cordova:2019bsd,Cordova:2019jqi,Apte:2022xtu,Brennan:2023kpo,Wan:2019oyr,Wan:2018djl,Wang:2017loc}. 
 
The anomalies that give these constraints can often be distinguished by evaluating the associated $(d+1)$-dimensional SPT phase on a mapping class torus. For example, for  $d$-dimensional, unitary continuum QFTs with certain types of discrete, abelian higher form global symmetries with an anomaly given by the SPT phase $\omega_{d+1}(A)\in H^{d+1}(BG;U(1))$, if 
\eq{\label{standardresult}
\exp\left\{2\pi i \oint_{N_{d+1}^{(g)}}\omega_{d+1}(A)\right\}\neq 1~,
}
where 
$N_{d+1}^{(g)}=(S^p\times S^{d-p})\rtimes_g S^1$ is the mapping class torus of $S^p\times S^{d-p}$ where $g\in G$ is a twist of the $G$-bundle on $S^p\times S^{d-p}$, then the theory cannot flow to a symmetry preserving gapped phase \cite{Cordova:2019bsd,Cordova:2019jqi}. The reason is that the variation of the partition function of any QFT under a constant (i.e. trivial) background gauge transformation with an anomaly leads to the relation 
\eq{
Z_{QFT}[A]=
\exp\left\{2\pi i \oint_{N_{d+1}^{(g)}}\omega_{d+1}(A)\right\}\times Z_{QFT}[A]~.
}
This implies that $Z_{QFT}[A]=0$ when the phase  \eqref{standardresult} is non-trivial, which cannot be accommodated by a unitary, symmetry preserving TQFT: $Z_{TQFT}[A]\neq0$. There also exists similar results for continuum QFTs with $Spin_G$-symmetries, time reversal symmetry, and particular types of non-invertible symmetries  \cite{Brennan:2023kpo,Apte:2022xtu,Wan:2019oyr,Wan:2018djl,Wang:2017loc}. 

\bigskip
In this paper, we will study the IR constraints that are imposed by a peculiar type of anomaly: one which does not correspond to a ``symmetry.'' Let us clarify what we mean by this. Consider a $d$-dimensional theory which has a (continuous) 0-form global symmetry group $G^{(0)}$ -- which we define by the group that acts effectively on gauge invariant local operators --  which has $\pi_1(G^{(0)})\neq 0$.  When we couple this symmetry to background gauge fields, the connection has a possible discrete flux $w_2(G^{(0)})\in H^2(BG^{(0)};\pi_1(G^{(0)}))$ which obstructs the $G^{(0)}$-connection from being able to be lifted to a $\widehat{G}^{(0)}$-connection. If we were to gauge $\hatG^{(0)}\supset G^{(0)}$, turning on this discrete flux would correspond to turning on a background gauge field for a 1-form center symmetry $\pi_1(G^{(0)})^{(1)}$ which acts on the $G^{(0)}$ Wilson lines. However, when $G^{(0)}$ connection is a background gauge field, the Wilson lines are trivial operators and the fluxes do not correspond to a symmetry because there is no corresponding charged operator. In this sense, we can say that the $w_2(G^{(0)})$ flux does not correspond to a ``symmetry'' (although they clearly are allowed background fluxes for $G^{(0)}$ symmetry).

In general, anomalies can depend on  this discrete flux $w_2(G^{(0)})$. For example, such anomalies may have a dependence \eq{\label{reduceanom}
\omega_{d+1}(A)=w_2(G^{(0)})\wedge \omega_{d-1}(A)~. 
}
This gives a class of anomalies that are only activated when we couple the theory to $G^{(0)}$-bundles with non-trivial discrete flux. It is in this sense that we mean that the anomaly can depend on particular background fluxes which are not associated to a symmetry. We will thus refer to these anomalies as \emph{background flux anomalies.}

%


In this paper, we will derive constraints in theories which have a higher form symmetry group
\eq{
G_{\rm global}=G_0\times G_{\rm disc.}\quad, \quad G_{\rm disc.}=G^{(0)}_{\rm disc.}\times G^{(p)}_{\rm disc.}~.
}
where $G_0$ is a continuous 0-form group with $\pi_1(G_0)\neq 0$ and $G_{\rm disc.}$ is a product of abelian discrete higher form global symmetry groups {with $p\neq 2$}. We will show that if the theory has anomalies that are only activated when $w_2(G_0)\neq 0$ of the form \eqref{reduceanom} which evaluates non-trivially on the mapping class torus:
\eq{
{\rm exp}\left\{2\pi i \oint_{N_{d+1}^{(g)}}w_2(G_0)\wedge \omega_{d-1}(A)\right\}\neq 1\quad, \quad N_{d+1}^{(g)}=(S^2\times S^{p+1})\rtimes_g S^1~,
}
then the theory cannot flow to a unitary, symmetry preserving, gapped phase {without symmetry fractionalization for $G_0$}.\footnote{To be clear, here we assume that there is no symmetry fractionalization in the UV theory {(which may occur when $p=1$)} and that any symmetry fractionalization is an emergent property in the IR.}

\bigskip
The outline of our paper is as follows. We will begin in Section \ref{sec:review} by reviewing some of the results from \cite{Cordova:2019bsd,Cordova:2019jqi} that proves the result \eqref{standardresult} for higher form discrete generalized global symmetries. Then in Section \ref{sec:new}, we will describe how these results can be generalized to flux background anomalies by using the clutching construction similar to the results of \cite{Brennan:2023kpo}. Then we will conclude in Section \ref{sec:examples} with a set of examples in $4d$ gauge theories. In $4d$, flux anomalies can be of one of the two forms:
\eq{
\CA_1=\frac{2\pi i }{N}\int w_2(G^{(0)})\cup x_1\cup y_2\quad, \quad \CA_2=\frac{2\pi i}{N}\int w_2(G^{(0)})\cup z_3~,
}
where here  $x_1,w_2(G^{(0)}),y_2,z_3\in H^\ast(N_{d+1};\IZ_N)$ with degree indicated by their subscript. We will give two interesting examples of theories that have these anomalies: $SU(2)$ QCD with a scalar field and $\CN=2$ $SU(2)$ gauge theory with $N_f>1$ flavors and comment on their generalization to other gauge groups and matter content. 

\section{Review of Anomaly Enforced Gaplessness for Discrete Symmetries}
\label{sec:review}

In this section we will review the results of \cite{Cordova:2019bsd,Cordova:2019jqi} on anomaly enforced gaplessness for theories with anomalous discrete higher form global symmetries. In this paper, the authors studied  $d$-dimensional QFTs that have an abelian discrete global symmetry which is given by a product of higher form global symmetries:
\eq{
G_{\rm global}=G^{(0)}\times G^{(p)}\times G^{(d-p-2)}~,
}
and have an anomaly that is described by the $(d+1)$-dimensional SPT phase $\omega_{d+1}(A)\in H^{d+1}(M;U(1))$ where $A$ collectively denotes the background gauge fields for $G_{\rm global}$. Their result is given as follows: \\

\noindent \textbf{Theorem} \cite{Cordova:2019bsd,Cordova:2019jqi}: A $d$-dimensional, unitary TQFT with a symmetry group $G_{\rm global}$ as defined above cannot carry an anomaly described by an SPT phase $\omega_{d+1}$ if 
\eq{
\exp\left\{2\pi i \oint_{N_{d+1}^{(g)}}\omega_{d+1}(A)\right\}\neq 1~, 
}
where $N_{d+1}^{(g)}$ is the mapping class torus where we twist the $G_{\rm global}$-bundle by $g\in G^{(0)}$. \\

In particular, this implies that any QFT with such an anomaly for this class of symmetry cannot flow to a symmetry preserving gapped phase: it must flow to either a gapless phase or a gapped phase that spontaneously breaks the $G_{\rm global}$ symmetry. 

\bigskip
\noindent This theorem can be proven as follows. 

\bigskip
\noindent\textit{pf.} Consider a $d$-dimensional unitary TQFT. Such theories are by definition reflection positive \cite{Freed:2016rqq}, which means that for any closed $d$-dimensional manifold $X_d$ which can be decomposed as $X_d=\overline{Y}_d\#Y_d$, the partition function is non-negative: $Z[X_d]\geq0$. Here we use the notation where $\overline{Y}_d$ is the orientation reversal of $Y_d$. This follows from the fact that the path integral on the manifold $Y_d$ defines a state in the Hilbert space defined on the boundary $Z[Y_d]:=|Y_d\rangle\in \CH[\partial Y_d]$ so that the path integral on $X_d$ defines the norm of $|Y_d\rangle$:
\eq{
Z[X_d]=\langle Y_d|Y_d\rangle\geq0~,
}
which is non-negative in any unitary theory. 

In particular, the fact that $S^d$ can be decomposed as $S^d=\overline{D}^d\#D^d$ implies $Z[S^d]\geq0$. However, since $D^d$ is contractible, it defines the vacuum state which in any unitary TQFT is non-trivial. This implies that with an appropriate choice of normalization{/counter-terms:}
\eq{
Z[S^d]=\langle D^d|D^d\rangle=1~.
}
This choice of normalization reflects the fact that in any unitary TQFT, the Hilbert space on $\partial D^d=S^{d-1}$ is generated by the vacuum state:
\eq{
\CH[S^{d-1}]\cong \IC\Big[|D^d\rangle\Big]~,
}
since there is a unique local operator (the identity operator).

The $d$-sphere also admits a decomposition 
\eq{
S^d=\left(\overline{D^{d-p}\times S^p}\right)\#\left(S^{d-p-1}\times D^{p+1}\right)~,
}
where we glue along a $S^{d-p-1}\times S^p$. Since $Z[S^d]=1$, the above decomposition implies 
\eq{
Z[S^d]=\langle D^{d-p}\times S^p|S^{d-p-1}\times D^{p+1}\rangle
=1~,
}
and therefore that the state $|D^{d-p}\times S^p\rangle$ is non-trivial, which by unitarity also implies 
\eq{{
\langle D^{d-p-1}\times S^{p+1}|D^{d-p-1}\times S^{p+1}\rangle=Z[S^{d-p-1}\times S^{p+1}]>0~,}
}
for any $p$. 

Now let us use the Lemma from the next subsection which states that a TQFT with a symmetry preserving vacuum has a non-vanishing partition function on $S^{d-p-1}\times S^{p+1}$ when we turn on background gauge fields $A$ for $G_{\rm global}$:
\eq{
Z[S^{d-p-1}\times S^{p+1};A]>0~. 
}
We can use this fact to derive a contradiction. We also need the fact that the anomalous variation of the partition function can be  given by evaluating the SPT phase on the mapping class torus with 
\eq{
Z[A^{g(x)}]=\exp\left\{ 2\pi i \oint_{N_{d+1}^{g(x)}}\omega_{d+1}(A)\right\}\,Z[A]~, 
}
where $A^{g(x)}$ is the gauge variation of $A$ with respect to $g(x)$ which is valued in $G^{(0)}${, and $N_{d+1}^{g(x)}=( S^{d-p-1}\times S^{p+1})\rtimes_{g(x)} S^1$ is the mapping class torus where the $G_{\rm global}$-bundle is twisted by the action of $g(x)\in G^{(0)}$}. In particular, when $g(x)=g$ is the constant function, $A^{g}=A$ since we have taken $G_{\rm global}$ to be abelian. However, even when $g(x)=g$ is the constant function (which trivially acts on the gauge fields), the anomalous phase can be non-zero as the twisting by $g$ turns on a flat $G^{(0)}$ gauge field along the $S^1$. The anomalous variation then reads 
\eq{
Z[A]=\exp\left\{ 2\pi i \oint_{N_{d+1}^{g}}\omega_{d+1}(A)\right\}\,Z[A]~, 
}
which implies 
\eq{
\text{1.) }\exp\left\{ 2\pi i \oint_{N_{d+1}^{g}}\omega_{d+1}(A)\right\}=1\quad {\rm or}\quad \text{2.) }Z[A]=0~.
}
Therefore, if $\exp\left\{2\pi i \oint_{N_{d+1}^{g}}\omega_{d+1}(A)\right\}\neq 1$, then $Z[A]=0$ and 
the anomaly cannot be a unitary, symmetry preserving TQFT as we have  $Z_{TQFT}[S^{p+1}\times S^{d-p-1};A]\neq 0$, which proves the theorem. \qed

\subsection{Proof of Lemma}

In order to complete our review, we need to prove the following lemma. 

\bigskip
\noindent\textbf{Lemma} \cite{Cordova:2019bsd,Cordova:2019jqi}: For a $d$-dimensional unitary, symmetry preserving TQFT with $G_{\rm global}=G^{(0)}\times G^{(p)}\times G^{(d-p-2)}$ a discrete, abelian group, then 
\eq{ Z[S^{d-p-1}\times S^{p+1};A]>0~.}

\bigskip
\noindent \textit{pf.} Here, the background gauge fields can be turned on by wrapping a $G^{(d-p-2)}$ symmetry defect operator $U_g$ on $S^{p+1}$ and a $G^{(p)}$ symmetry defect operator  $\CU_{g^\prime}$ on $S^{d-p-1}$. 
Let us first consider the simple case where $g^\prime=1\in G^{(p)}$. Now, we can relate
\eq{
Z[S^{p+1}\times S^{d-p-1};A]=\langle U_g(S^{p+1})\rangle ~.
}
In this case, we can cut the space time along a great circle of $S^{d-p-1}$:\footnote{In our discussion here we assume that the symmetry defect operators to be inserted along embedded manifolds which are away from any topological manipulations unless otherwise specified.}
\eq{
\langle U_g(S^{p+1})\rangle =\langle S^{p+1}\times D^{d-p-1}|U_g^{(p+1)}\rangle ~,
}
where the state $|U_g^{(p+1)}\rangle$ is the path integral on $S^{p+1}\times D^{d-p-1}$ where $U_g$ is wrapped on $S^{p+1}$. Since the state $|U_g^{(p+1)}\rangle$ is an element of the Hilbert space $\CH[S^{p+1}\times S^{d-p-2}]$, we can compute its inner product with the state $|D^{p+2}\times S^{d-p-2}\rangle$:
\eq{
\langle D^{p+2}\times S^{d-p-2}|U_g^{(p+1)}\rangle=\langle U_g(S^{p+1})\rangle_{S^d}=Z[S^d]=1~. 
}
Further, if $G^{(d-p-2)}$ is preserved then $U_g(S^{p+1})$  has trivial winding with all operators on $S^{p+1}\times S^{d-p-1}$ and $|U_g^{(p+1)}\rangle=|S^{p+1}\times D^{d-p-1}\rangle$. {This follows if the symmetry is preserved because then any charged operator necessarily vanishing expectation value.} This implies that 
\eq{\label{lemma1}
\langle U_g(S^{p+1})\rangle_{S^{p+1}\times S^{d-p-1}}&=\langle U_g^{(p+1)}|S^{p+1}\times D^{d-p-1}\rangle\\
&=\langle S^{p+1}\times D^{d-p-1}|S^{p+1}\times D^{d-p-1}\rangle>0~. 
}

Now consider the case where $g^\prime\neq 1$. Now if we cut along the same great circle in $S^{d-p-1}$, we get a state $|\phi_g^\prime\rangle\in \CH[S^{p+1}\times S_{g^\prime}^{d-p-2}]$ in the defect (i.e. $g^\prime$-twisted) Hilbert space:
\eq{
\langle U_g(S^{p+1})\,\CU_{g^\prime}(S^{d-p-1})\rangle_{S^{p+1}\times S^{d-p-1}}=\langle \phi^\prime_1|\phi^\prime_g\rangle~,
}
where $|\phi^\prime_g\rangle$ is defined by the path integral on $S^{p+1}\times D^{d-p-1}$ with $U_g$ wrapping a $S^{p+1}$ and $\CU_{g^\prime}$ wrapping $D^{d-p-1}$. 

If there exists an invertible linear map $f:\CH[S^{p+1}\times S^{d-p-2}_{g^\prime}]\to \CH_{g^\prime}[S^{d-1}]$, where $\CH_{g^\prime}[S^{d-1}]$ is the defect/twisted Hilbert space where $\CU_{g^\prime}$ wraps a $S^{d-p-2}$ circle on the boundary, such that 
\eq{
f:|\phi^\prime_g\rangle\longmapsto |0_{g^\prime}\rangle~, 
}
then the fact that $\langle 0_g|0_g\rangle=1$ implies that $\langle \phi^\prime_g|\phi^\prime_g\rangle\neq 0$. This map can be constructed as in \cite{Cordova:2019bsd,Cordova:2019jqi} by taking $D^d\backslash (D^{p+2}\times S^{d-p-2})$ where $\CU_g$ wraps a cylinder $S^{d-p-2}\times [0,1]$ running from one boundary component to the other. 
 Since $f$ is manifestly linear, we simply need to show that it is not the trivial map. This can be shown by computing 
\eq{
\langle 0_{g^\prime}|f|\phi^\prime_1\rangle=\langle \CU_{g^\prime}(S^{d-p-1})\rangle_{S^{p+1}\times S^{d-p-1}}>0~, 
} 
by equation \eqref{lemma1}. Here we used the fact that $\langle 0|f=\langle S^d\backslash(D^{p+2}\times S^{d-p-2})|$ and that 
\eq{
\Big( S^d\backslash(D^{p+2}\times S^{d-p-2})\Big)\#(S^{p+1}\times D^{d-p-1})&\cong(S^{p+1}\times D^{d-p-1})\#(S^{p+1}\times D^{d-p-1})\\
&= S^{p+1}\times S^{d-p-1}~. }

We can now implement state-operator correspondence to show that the state $|\phi^\prime_1\rangle$ corresponds to a $(d-p-2)$-dimensional operator along which $\CU_{g^\prime}$ can end. It also follows that there is a similar operator along which $U_g$ can end. 

Now if we consider the partition function then we can cut the $U_g$ and $\CU_{g^\prime}$ by the insertion of the operator $\phi$ and $\phi^\prime$ respectively:
\eq{
Z[S^{p+1}\times S^{d-p-1};A]&=\langle U_g(S^{p+1})\,\CU_{g^\prime}(S^{d-p-1})\rangle_{S^{p+1}\times S^{d-p-1}}\\
&=\langle U_g^{\phi}(D^{p+1})\,\CU^{\phi^\prime}_{g^\prime}(D^{d-p-1})\rangle_{S^{p+1}\times S^{d-p-1}}~,
}
where $U_g^{\phi}(D^{p+1})$ and $\CU_{g^\prime}^{\phi^\prime}(D^{d-p-1})$ are the symmetry defect operators inserted on the disk with an operator $\phi,\phi^\prime$ inserted along the boundary. 

We can then contract the operators to a point operator using topological invariance. However, since there is a unique local operator (the identity operator), we see that they must contract to the identity operator. This means that the partition function is given 
\eq{
Z[S^{p+1}\times S^{d-p-1};A]&=\langle U_g(S^{p+1})\,\CU_{g^\prime}(S^{d-p-1})\rangle_{S^{p+1}\times S^{d-p-1}}\\
&=\langle U_g^{\phi}(D^{p+1})\,\CU^{\phi^\prime}_{g^\prime}(D^{d-p-1})\rangle_{S^{p+1}\times S^{d-p-1}}\\&
=\langle \mathds{1}\rangle_{S^{p+1}\times S^{d-p-1}}=Z[S^{p+1}\times S^{d-p-1};0]>0~. 
}
This completes the proof. See \cite{Cordova:2019bsd,Cordova:2019jqi} for more details and further discussion.  \qed

\section{Anomaly Enforced Gaplessness for Background Flux Anomalies}
\label{sec:new}

In this section we would like to consider the constraints imposed by a particular class of anomalies: background flux anomalies. As described in the introduction, these are in a sense anomalies for which we turn on background fields that do not correspond to any symmetry. 

Our setting is a $d$-dimensional unitary quantum field theory with a global symmetry group $G_{\rm global}$ which has the form 
\eq{\label{Gglobal}
G_{\rm global}=G_0\times G_{disc.}=G_0\times\left(G^{(0)}\times G^{(p)}\right)~,
}
where $d=p+3$ and $p\neq 2$. 
Here $G_0$ is a continuous 0-form symmetry group with $\pi_1(G_0)\neq 0$ and $G_{disc.}$ is a product of discrete, abelian higher form symmetry groups. We assume here that the UV theory has no innate symmetry fractionalization in the case where $p=1$. 

We would like to make a comment about the fact that $\pi_1(G_0)\neq 0$. If we were to gauge $G_0$, then $\pi_1(G_0)$ would correspond to a 1-form magnetic global symmetry. However, since $G_0$ is a global symmetry, there are no intrinsic $G_0$ line operators that would be charged under a putative symmetry. Rather, the physical property associated to $\pi_1(G_0)$ corresponds to charge fractionalization. 

 As we discussed in the introduction, $G_0$-symmetry fractionalization occurs when a line operator transforms projectively under $G_0$.  Such a line operator exhibits symmetry fractionalization if it has a world volume anomaly which 
 is given by a non-trivial element of the cohomology group $H^2(BG_0;\pi_1(G_0))$ which correspond to obstruction classes for lifting $G_0$-bundles to $\hatG_0$-bundles. 
We often denote these characteristic classes as $w_2(G_0)\in H^2(BG_0;\pi_1(G_0))$.

 It would also be consistent to consider a theory with $\hatG_0$ global symmetry, it would simply be that $\pi_1(G_0)\subset \hatG_0$ does not act on any gauge invariant local operators. The choice of $G_0$ versus $\hatG_0$ global symmetry is a statement about what background gauge bundles we are allowed to couple to as well as what counter terms we allow in our theory. In particular, in the theory with $G_0$ symmetry, we are only allowed counter terms that are dependent on $G_0$ background connections. This means that in the theory with $G_0$ symmetry it is meaningful to measure $G_0$-symmetry fractionalization whereas this is not meaningful in the theory with $\hatG_0$ global symmetry.

In this paper, we are interested in generalizing the discussion from the previous section to anomalies that can only be activated by $G_0$-bundles which do not lift to $\hatG_0$-bundles. These are anomalies which are controlled by a $(d+1)$-dimensional SPT phase $\omega_{d+1}[A]\in H^{d+1}(BG_{\rm global};U(1))$ that vanishes on $G_0$-bundles which can be lifted to $\hatG_0$-bundles. In other words, they depend explicitly on the obstruction class $w_2(G_0)\in H^2(BG_0;\pi_1(G_0))$: $\omega_{d+1}[A;w_2(G_0)]$ with $\omega_{d+1}[A;0]=0$. We refer to these anomalies as ``background flux anomalies'' as they are anomalies that are only activated when we turn on particular background fluxes. 

Again, we want to be able to constrain IR phases by using the mapping class torus. To do so, we need to prove that 
\eq{
Z\left[S^2\times S^{p+1};A\right]\neq 0~,
}
for a unitary, symmetry preserving TQFT. In particular, we need to show that coupling a TQFT to a $G_0$-bundle with a non-trivial obstruction class $w_2(G_0)$ does not cause the partition function to vanish.

\subsection{Flux Backgrounds and the Clutching Construction}

In order to prove that coupling a TQFT to a non-trivial $G_0$-bundle which does not lift to a $\hatG_0$-bundle does not cause the partition to vanish, we need to make use of the clutching construction. 

Let us consider a $G_0$-bundle $P$ over a 2-sphere $S^2$. The 2-sphere can be covered by two overlapping open discs $D^2_{N/S}$ which cover the northern and southern hemisphere respectively. For  generic $P$, the restriction $P\big{|}_{D^2_{N/S}}$ to $D^2_{N/S}$ can be trivialized separately on each patch. The topology of the $G_0$-bundle is then specified by the transition function on the annulus $D^2_N\cap D^2_S$. These transition functions are classified by $\pi_1(G^{(0)})$, which specify the topological class of the bundle. 

For example, if  $G_0=U(1)$, then the bundle $P$ over $S^2$ will be specified by an element $n\in \pi_1(G_0)\cong \IZ$. The clutching construction above is the standard construction of the holomorphic line bundle of degree $n$ over $S^2\cong \ICP^1$, $P=\CO(n)$, where $n$ specifies the first Chern class $n=\oint\frac{F_{U(1)}}{2\pi}$. 

For more general $G_0$, the choice of element $[f]\in \pi_1(G_0)$ specifies a discrete flux $[f]\mapsto w_2(G_0)\in H^2(BG_0; \pi_1(G_0))$ which can be seen explicitly by restricting to a $U(1)\subset G_0$ for which the particular element $[f]\in \pi_1(G_0)$ can be represented by an element $[\overline{f}]\in \pi_1(U(1))$ and noting that the corresponding $\frac{F_{U(1)}}{2\pi} $ is an integer lift of $w_2(G_0)$. 

We can now use this to derive under what conditions $Z[S^2\times S^{p+1};w_2(G_0)]\neq 0$. Here we will take a construction which is similar to the Gluck twist as in \cite{Glu62,Reutter:2020bav,Brennan:2023kpo}. Since a $G_0$-bundle on $S^{2}\times S^{p+1}$ with non-trivial $w_2(G_0)$ on the first factor can be constructed by the twisted gluing 
\eq{
\left(\overline{D^2\times S^{p+1}}\right)\#_f \left(D^2\times S^{p+1}\right)~, 
}
where $f\in \pi_1(G_0)$, in the TQFT we can write 
\eq{
Z\left[S^2\times S^{p+1};w_2(G_0)\right]=\left\langle D^2\times S^{p+1}\Big{|}\widehat\CT_f\Big{|}D^2\times S^{p+1}\right\rangle~, 
} 
where here $\widehat\CT_f$ is a twist operator and $w_2(G_0)$ is specified by $[f]$ as described above. This operator enacts the twist of the $G_0$-bundle by {shifting the transition functions by} the element $f$. This is similar to the Gluck twist operator which twists the space/tangent bundle by the non-trivial element $F\in \pi_1(SO(4))$ to turn $(D^2\times S^2)\#_F(D^2\times S^2)=\overline{\ICP}^2\#\ICP^2$ into a non-spin manifold \cite{Glu62,Reutter:2020bav,Brennan:2023kpo}. 

Now, it is clear that if $\widehat\CT_f$ is the trivial operator in our theory, then 
\eq{
Z\left[S^2\times S^{p+1};w_2(G_0)\right]=Z\left[S^2\times S^{p+1};0\right]\neq 0~,
}
which would achieve our goal. We now need to physically interpret what it means for $\widehat\CT_f$ to be the trivial operator. 

Here, the twist operator $\widehat\CT_f$ is acting on the Hilbert space $\CH\left[S^1\times S^{p+1}\right]$. This Hilbert space is generated by local operators and, when $p=1$, line operators that wrap the $S^1$. In order for an operator to be charged under the twist operator, it must transform projectively under $G_0$ since $f$ forms a closed loop in $G_0$. By  assumption, there are no local operators which transform projectively under $G_0$. However, as discussed in the introduction, when $p=1$, there will exist some line operator (wrapping $S^1$) that will the generate a state that transforms projectively under $G_0$ iff the $G_0$ symmetry is {fractionalized} in the IR.\footnote{{
Note that the assumption that a TQFT has a group-like symmetry given by \eqref{Gglobal}  necessarily implies that the theory only has operators of co-dimension $1,~(p+1)$ (symmetry defect operators) and possibly operators of dimension $0,p$ that are charged under the corresponding symmetries. }
} 


Thus, we find that in a unitary theory without $G_0$-symmetry fractionalization (which only exists for $p=1$) that we have 
\eq{
Z\left[S^2\times S^{p+1};w_2(G_0)\right]=Z[S^2\times  S^{p+1};0]~,
}
which implies 
\eq{
Z\left[S^2\times S^{p+1};w_2(G_0)\right]\neq 0~. 
}

\subsection{Constraints from Mapping Class Tori}

Now we can derive constraints on IR phases that are imposed by background flux anomalies. Consider a $d$-dimensional unitary QFT with a symmetry group $G_{\rm global}$ of the form in \eqref{Gglobal} and let us assume that the theory only has a  background flux anomaly given by $\omega_{d+1}[A;w_2(G_0)]$. Let us take the partition function on the manifold $S^2\times S^{p+1}$. 
\eq{
Z_{QFT}[S^2\times S^{p+1};A]~.  
}

Given that the partition function of a TQFT does not vanish when we couple the theory to a $G_0$-bundle with a non-trivial obstruction class $w_2(G_0)$, we can consider the variation of the partition function with respect to $G^{(0)}$ background gauge transformations:
\eq{
Z_{QFT}[S^2\times S^{p+1};A^{g(x)}]=&\exp\left\{i \oint_{N_{d+1}^{g(x)}} \omega_{d+1}[A;w_2(G_0)]\right\}\times Z_{QFT}[S^2\times  S^{p+1};A]~,
}
where again $A^{g(x)}$ is the gauge transformation of $A$ with respect to $g(x)\in G^{(0)}$ and $N_{d+1}^{g(x)}=( S^2\times S^{p+1})\rtimes_{g(x)} S^1$ is the mapping class torus where the $G_{\rm global}$-bundle is twisted by the action of $g(x)\in G^{(0)}$. Again, in the case where $g(x)=g$ is the constant function, $A^{g(x)}=A$ and the equation becomes 
\eq{\label{QFTanomvar}
Z_{QFT}[S^2\times S^{p+1};A]=&\exp\left\{i \oint_{N_{d+1}^{g}} \omega_{d+1}[A;w_2(G_0)]\right\}\times Z_{QFT}[S^2\times  S^{p+1};A]~,
}
which implies that either 
\eq{
\text{1.) }\exp\left\{i \oint_{N_{d+1}^{(g)}} \omega_{d+1}[A;w_2(G_0)]\right\}=1\quad, \quad \text{2.) }Z_{QFT}[S^2\times  S^{p+1};A]=0~.
}
This implies that if the SPT phase evaluated on the mapping class torus is non-trivial, then the partition function must vanish. 

However, if the QFT flows to a gapped phase, the anomaly -- and therefore the anomalous variation \eqref{QFTanomvar} -- must also be matched by a TQFT. As we have shown for a unitary, symmetry preserving TQFT without charge fractionalization 
\eq{
Z_{TQFT}[S^2\times  S^{p+1};A]\neq 0~.
}
This implies that if the SPT phase evaluated on the mapping class torus is non-trivial, then the theory cannot flow to a gapped phase that preserves the symmetry and, when $p=1$, has no symmetry fractionalization. This proves our main result.

\section{Examples in $4d$}
\label{sec:examples}

Let us now consider some examples of theories with background flux anomalies. Here we will focus on $4d$ gauge theories. Because of dimensionality, there are only two types of possible flux anomalies. Here we can take $G_{\rm global}=G_0\times (G^{(0)}\times G^{(1)})$ or $G_{\rm global}=G_0\times G^{(1)}$ and have anomalies 
\eq{
\CA_1=\frac{2\pi i n}{N}\int w_2(G_0)\cup x_1\cup y_2\quad, \quad \CA_2=\frac{2\pi i n}{N}\int w_2(G_0)\cup Bock[y_2]~, 
}
where $x_1,y_2$ are gauge fields for $G^{(0)},G^{(1)}$ -- {which for simplicity we take to both be $\IZ_N$ and $Bock$ is the associated Bockstein map}. In these examples, the $G_0$  symmetry can fractionalize when line operators charged under $G^{(1)}$ transform projectively under $G_0$. 

Our results can be applied can be applied to anomalies of the type $\CA_1,\CA_2$ as they evaluate non-trivially on $(S^2\times S^2)\rtimes_g S^1$. 
Our results imply that theories with these anomalies cannot flow to a gapped phase without spontaneous symmetry breaking or symmetry fractionalization. Examples of TQFTs that match the $\CA_1,\CA_2$ anomaly without spontaneous symmetry breaking are given by 
\eq{
S_1&=\frac{2\pi i }{N}\int \left(b_2\cup \delta a_1+n \,b_2\cup w_2(G_0)+a_1\cup x_1\cup y_2\right)~,\\
S_2&=\frac{2\pi i }{N}\int \left(b_2\cup \delta a_1+n \,b_2\cup w_2(G_0)+a_1\cup Bock[y_2]\right)~,
}
respectively. Both of these theories, are described by a $\IZ_N$ gauge theory where the Wilson line has a world-volume anomaly $n\, w_2(G_0)$. 

We will now exhibit some theories that have this anomaly. 

\subsection{$4d$ $SU(2)$ QCD}
\label{sec:nonSUSY} 

Consider $4d$ $SU(2)$ gauge theory with $2N_f$ fermions in the fundamental representation $\psi^A$, $2N_f$ scalar fields in the fundamental representation $h^A$, as well as a real scalar in the adjoint representation $\Phi$ and an uncharged fermion $\chi$. Consider the interactions:
\eq{\label{nonSUSYaction}
S=\int d^4x \Big(&\frac{1}{2g^2}\Tr[F^2]+i \overline\psi_A\slashed{D}\psi^A+i \overline\chi\slashed{\partial}\chi+|Dh^A|^2+\Tr[(D\Phi)^2]\\
&-V(\Phi,h^A)+i \lambda_1\, \chi\psi^A\overline{h}_A+i\lambda_2 \delta_{AB} \,\psi^A\Phi \psi^B+c.c. \Big)~.
} 
Here the two Yukawa interactions break the global symmetry group down to $SO(2N_f)\times U(1)$ where the fields transform as

\begin{center}
\begin{tabular}{c|ccc}
&$SU(2)$&$SO(2N_f)$&$U(1)$\\
\hline
$\psi^A$&$\mathbf{2}$&$\mathbf{2N_f}$&0\\
$h^A$&$\mathbf{2}$&$\mathbf{2N_f}$&1\\
$\Phi$&$\mathbf{3}$&$\mathbf{1}$&0\\
$\chi$&$\mathbf{1}$&$\mathbf{1}$&$1$
\end{tabular}
\end{center}

Here we would like to comment on two features of these global symmetries. First, note that $U(1)$ has a cubic anomaly, which would imply that the theory must flow to a gapless theory. However, we can additionally break the $U(1)$ global symmetry by introducing a mass term for $\chi$ so that there is only an $SO(2N_f)$ global symmetry which has no perturbative anomalies.

More importantly, notice that  the center of $SO(2N_f)$ is gauged since it is identified with $-\mathds{1}_{SU(2)}\sim -\mathds{1}_{SO(2N_f)}$. This implies that there is an additional $\IZ_2$ flux background that can be activated. In the case where $N_f=2n_f$ the total background fluxes form a $\IZ_2\times \IZ_2$ generated by $w_2^{(L)},w_2^{(R)}\in H^2(PSO(2N_f);\IZ_2)$ whereas when $N_f=2n_f+1$, the total background flux forms a $\IZ_4$ which is generated by $w_2(\IZ_4)\in H^2(PSO(2N_f);\IZ_4)$. 

Besides being a $4d$ interacting QCD-like theory, this theory is of physical interest because the Yukawa interaction $(\lambda_1)$ is reminiscent of the Yukawa interaction in the Standard model. Indeed, if we choose the scalar potential $V$ such that the Higgs field $h^A$ condenses, then the fermion fields will also partially gain a mass. 

Now we would like to demonstrate the fact that this theory has an anomaly of the form $\CA_2$. The simplest way to see this is to compute the anomaly in the $U(1)$ Maxwell phase that results from condensing $\Phi$. As discussed in \cite{Brennan:2022tyl}, an anomaly involving background fluxes for $SO(2N_f)$ can be computed in the Maxwell phase by computing the symmetry fractionalization of the emergent $U(1)^{(1)}_e\times U(1)^{(1)}_m$ global symmetries. 
In the Maxwell phase, all fundamental fermions are integrated out and the fractionalization of the emergent $U(1)^{(1)}_e$ is given by the charge of the fundamental fermion under $PSO(2N_f)$. This means that turning on the background flux for $PSO(2N_f)$ results in 
\eq{\label{B2electric}
\frac{B_2^{(e)}}{2\pi}=\begin{cases}
\half(w_2^{(L)}+w_2^{(R)})& N_f~\text{even}\\
\half w_2(\IZ_4)& N_f~\text{odd}
\end{cases}
}
which is the obstruction for lifting $PSO(2N_f)\to SO(2N_f)$ since $\psi^A$ transforms under the vector representation of $SO(2N_f)$. 

The background flux for $B_2^{(m)}$ then corresponds to the projective representation for the Hilbert space of fermion zero-modes bound to the minimal smooth monopole. Using the Callias Index theorem \cite{Callias:1977kg}, we see that there are $2N_f$ real fermion zero-modes so that the monopole Hilbert space transforms as a Dirac spinor of $PSO(2N_f)$. This means that turning on the background flux for $PSO(2N_f)$ results in 
\eq{\label{B2magnetic}
\frac{B_2^{(m)}}{2\pi}=\begin{cases}
\half w_2^{(L)} & N_f~\text{even}\\
\frac{1}{4} w_2(\IZ_4)& N_f~\text{odd}
\end{cases}
}
which is the obstruction for lifting $PSO(2N_f)\to Spin(2N_f)$ since the Hilbert space transforms in the spin representation of $SO(2N_f)$. 

After implementing the Wu relations, we find that on spin manifolds, that the anomaly of the UV theory has is given by 
 \eq{
 \CA
 =\begin{cases}
\pi i \int w_2^{(L)}\cup w_3^{(R)}&N_f~\text{even}\\
\pi i  \int w_2(\IZ_4)\cup w_3(\IZ_4)&N_f~\text{odd}
 \end{cases}
 }
Thus we find that that the anomaly can only be activated on spin manifolds -- and in particular can be activated on the mapping class torus -- allowing us to apply our theorem. \\


\bigskip
Notice here that the anomaly we have derived is only dependent on the background fluxes. So it is prudent to ask in what sense this is an anomaly at all. The key is that the additional flux background that arises from the identification of $-\mathds{1}_{SU(2)}\sim -\mathds{1}_{SO(2N_f)}$ can be identified as a (0,1)-form symmetry or a ``color-flavor center'' symmetry \cite{Brennan:2023tae,Cherman:2017tey}. In fact, since our theorem can be applied to this background, it is a proof by example that these symmetries are physical: they can have anomalies and that their anomalies obstruct the theory from flowing to a trivially gapped phase.

\subsection{$4d$ $\CN=2$ $SU(2)$ QCD with $N_f>1$ Hypermultiplets}
 \label{sec:SUSY}

 Now consider $4d$ $\CN=2$ $SU(2)$ QCD with $N_f$ hypermultiplets. In addition to the gauge field, the matter content of this theory is given by\footnote{Note that here $\Phi$ is a complex scalar field as opposed to the previous example where it was a real-valued scalar field.} 
 
\begin{center}
\begin{tabular}{c|cccc}
&$SU(2)$&$SO(2N_f)$&$SU(2)_R$&$U(1)_R$\\
\hline
$\psi^A$&$\mathbf{2}$&$\mathbf{2N_f}$&$\mathbf{1}$&-1\\
$h^A$&$\mathbf{2}$&$\mathbf{2N_f}$&$\mathbf{2}$&0\\
$\Phi$&$\mathbf{3}$&$\mathbf{1}$&$\mathbf{1}$&2\\
$\lambda$&$\mathbf{3}$&$\mathbf{1}$&$\mathbf{2}$&$1$
\end{tabular}
\end{center}

Here, the $U(1)_R\to \IZ_r$ where $r=|8-2N_f|$ by an ABJ anomaly when $N_f\neq 4$ and additionally $4\geq N_f\geq 0$ for asymptotically free theories. The resulting $\IZ_r$ symmetry has a cubic anomaly 
\eq{
[\IZ_r]^3=6-4N_f~{\rm mod}_r~,
}
as well as a mixed anomaly with $SO(2N_f)$:
\eq{
[\IZ_r]\times[SO(2N_f)]^2=-2~{\rm mod}_r~. 
}
Note that both of these `t Hooft anomalies vanish in the case where $N_f=3$ since $r=2$.

In addition to the standard kinetic terms for a gauge theory, the theory also has the additional interactions:
\eq{
\CL_{int}=&g\delta_{AB}\psi^A\Phi\psi^B+g^2[\Phi^\dagger,\Phi]^2+g^2\overline{h}_{Aa}\{\Phi^\dagger,\Phi\}h^{Aa}+g\psi^A\lambda^a\overline{h}_{Aa}\\
&+g^2(\overline{h}_{Aa}(\sigma^r)^a_{~b}h^{Ab})^2+c.c.
}
where here $A=1,..,2N_f$ and $a=1,2$ are $SO(2N_f)$ and $SU(2)_R$ indices respectively. Here the Yukawa terms again serve to break the $SU(2N_f)\times SU(2N_f)$ global symmetry of the interacting theory to a $SO(2N_f)$ symmetry. This theory has the same matter content transforming under the $SO(2N_f)$ symmetry (again, the gauge symmetry gauges $\IZ_2\subset SO(2N_f)$). We thus find again that the theory has a background flux anomaly\footnote{Note that we cannot apply Wu identities to trivialize the $N_f$-odd anomaly since the fluxes are $\IZ_4$-valued.}
\eq{
\CA=\begin{cases}
\pi i \int w_2^{(L)}\cup w_3^{(R)}&N_f\text{ even}\\
\pi i \int w_2(\IZ_4)\cup w_3(\IZ_4)&N_f\text{ odd}
\end{cases}
}
It is known that these theories flow to gappless theories in the IR via the work of Seiberg-Witten \cite{Seiberg:1994rs,Seiberg:1994aj}, thereby matching our theorem. 

\subsection{ SUSY and non-SUSY $Sp(N_c)$ QCD}

More generally, the flux background anomalies for $SU(2)$ QCD from Section \ref{sec:nonSUSY} and $\CN=2$ $SU(2)$ QCD in Section \ref{sec:SUSY} 
generalizes straightforwardly to $Sp(N_c)$ gauge groups. The reason is that the structure of the interactions are the same as above due to reality/pseudo-reality of the representations. 

For $Sp(N_c)$ non-SUSY QCD, we consider the theory with the action given in \eqref{nonSUSYaction} where the fields transform under $SO(2N_f)\times U(1)$ as\footnote{Again we can consider adding a mass term for $\chi$ to break the anomalous $U(1)$ global symmetry. }

\begin{center}
\begin{tabular}{c|ccc}
&$Sp(N_c)$&$SO(2N_f)$&$U(1)$\\
\hline
$\psi^A$&$\mathbf{2N_c}$&$\mathbf{2N_f}$&0\\
$h^A$&$\mathbf{2N_c}$&$\mathbf{2N_f}$&1\\
$\Phi$&$\mathbf{adj}$&$\mathbf{1}$&0\\
$\chi$&$\mathbf{1}$&$\mathbf{1}$&$1$
\end{tabular}
\end{center}

\noindent As before, we can flow to the Maxwell phase by condensing $\Phi$ and look for anomalies by implementing the fractionalization technique. Here condensing $\Phi$ leads to the breaking $Sp(N_c)\mapsto U(1)^{N_c}$. Due to the fact that the fermions transform under the same global symmetry group, we find that the allowed background fluxes are the same as before and that the background flux for $PSO(2N_f)$. 
Since $-\mathds{1}_{Sp(N_c)}\sim -\mathds{1}_{SO(2N_f)}$, the obstruction class from lifting $PSO(2N_f)\to SO(2N_f)$ will activate a $\IZ_2\subset (U(1)^{(1)}_e)^{N_c}$ discrete flux in the IR associated with the center $Z(Sp(N_c))$:
%
\eq{\label{B2electricSp}
\frac{B_2^{(e)}}{2\pi}=\begin{cases}
\half(w_2^{(L)}+w_2^{(R)})& N_f~\text{even}\\
\half w_2(\IZ_4)& N_f~\text{odd}
\end{cases}
}
Here $B_2^{(e)}$ is the background gauge field of a 1-form center symmetry for a choice of $U(1)_0\subset U(1)^{N_c}$ where $U(1)_0\supset Z(Sp(N_c))$.\footnote{Note that this flux will be independent of the choice of $U(1)_0$ because any two choices will have the same $\IZ_2\subset U(1)_0$. Said differently, two choices for $U(1)_0$ have $\IZ_2$-valued background gauge fields which have different integer lifts in $U(1)^{N_c}$. } 
 Here we can pick the generator of $U(1)_0$ to be $Q_0$ which acts as 
\eq{
Q_0=\left(\begin{array}{cc}
\mathds{1}_{N_c}&0\\
0&-\mathds{1}_{N_c}
\end{array}\right)~,
}
in the fundamental representation. 


To compute the background flux anomaly, 
we only need to check the mixed anomaly of the 1-form global symmetries of this $U(1)_0$. We then need to determine how the UV background fluxes activate the background gauge field $B_2^{(m)}$ for the $U(1)^{(1)}_m$ magnetic 1-form symmetry associated to $U(1)_0$. 
Now, since the $U(1)_0$ minimal monopole is the direct sum of $N_c$ $Sp(N_c)$ minimal monopoles, there will be $2N_cN_f$ real fermion zero-modes and turning on the background flux for $PSO(2N_f)$ will result in an IR background gauge field:
\eq{\label{B2magneticSp}
\frac{B_2^{(m)}}{2\pi}=\begin{cases}
\frac{N_c}{2}w_2^{(L)} & N_f\text{ even}\\
\frac{N_c}{4} w_2(\IZ_4)& N_f~\text{odd}
\end{cases}
}
We then find that that the anomaly of the UV theory is identical to that of $SU(2)$ QCD: 
 \eq{\label{SpNc}
 \CA
 =\begin{cases}
\pi i\,N_c \int w_2^{(L)}\cup w_3^{(R)}&N_f\text{ even}\\
\pi i\, N_c  \int w_2(\IZ_4)\cup w_3(\IZ_4)&N_f~\text{odd}
 \end{cases}
 }

For $\CN=2$ $Sp(N_c)$ QCD with $N_f$ hypermultiplets, the matter of the theory transforms under $SO(2N_f)\times SU(2)_R\times U(1)_r$ as: 

\begin{center}
\begin{tabular}{c|cccc}
&$Sp(N_c)$&$SO(2N_f)$&$SU(2)_R$&$U(1)_r$\\
\hline
$\psi^A$&$\mathbf{2N_c}$&$\mathbf{2N_f}$&$\mathbf{1}$&-1\\
$h^A$&$\mathbf{2N_c}$&$\mathbf{2N_f}$&$\mathbf{2}$&0\\
$\Phi$&$\mathbf{adj}$&$\mathbf{1}$&$\mathbf{1}$&2\\
$\lambda$&$\mathbf{adj}$&$\mathbf{1}$&$\mathbf{2}$&$1$
\end{tabular}
\end{center}

\noindent Now the $U(1)_r$ global symmetry is broken to $\IZ_r$ with $r=|8 N_c - 2 N_f|$ by an ABJ anomaly when $N_f\neq 4N_c$. Again, we find that $4N_c\geq N_f$ for asymptotically free theories. The $\IZ_r$ global symmetry also has a cubic anomaly and mixed `t Hooft anomaly with $SO(2N_f)$ which are given by the coefficients:
\eq{
[\IZ_r]^3=N_c (4 N_c + 4 - 2 N_f)~{\rm mod}_r\quad, \quad [\IZ_r]\times[SO(2N_f)]^2=-2N_c~{\rm mod}_r~. 
}
Now, when $N_f=3N_c$ the theory perturbative anomalies all vanish. 

However, as in the case of $\CN=2$ $SU(2)$ QCD, this theory still has a non-perturbative background flux anomaly. As above, the computation of the flux anomaly follows straightforwardly from the non-supersymmetric case: the minimal IR Wilson line has charges \eqref{B2electricSp} under $PSO(2N_f)$ and the minimal monopole has $2N_f$ zero-modes so that the corresponding fluxes for $B_2^{(m)}$ are given by \eqref{B2magneticSp}. Therefore we the theory has the anomaly 
\eq{\label{SpNcN2}
\CA=\begin{cases}
\pi i\, N_c \int w_2^{(L)}\cup w_3^{(R)}&N_f\text{ even}\\
\pi i\, N_c\int w_2(\IZ_4)\cup w_3(\IZ_4)&N_f\text{ odd}
\end{cases}
}
Thus we find that non-SUSY $Sp(N_c)$ QCD when $N_f$ is even and  $\CN=2$ $Sp(N_c)$ QCD have a non-trivial anomaly \eqref{SpNc} and \eqref{SpNcN2} respectively which can be activated on the mapping class torus when $N_c$ is odd, allowing us to apply our theorem.  And indeed, while the IR phase of non-SUSY QCD is generally unknown, the $\CN=2$ SUSY theories are also known to flow to gapless theories similar to the $\CN=2$ $SU(2)$ QCD theory above \cite{Argyres:1995fw,DHoker:1996kdj}, thereby satisfying our theorem.

\section*{Acknowledgements}
The authors would like to thank Ken Intriligator,  Po-Shen Hsin, Kantaro Ohmori, Clay C\'ordova and Thomas Dumitrescu for helpful discussions and related collaborations. 
TDB is supported by Simons Foundation award 568420 (Simons
Investigator) and award 888994 (The Simons Collaboration on Global Categorical Symmetries).

\bibliographystyle{utphys}
\bibliography{FluxAnomaliesBib}

\end{document}